\documentclass[10pt,floatfix,aps,nofootinbib,superscriptaddress,twocolumn]{revtex4} 

\usepackage{amssymb,amsfonts,multirow}

\usepackage{epsfig}

\usepackage{placeins}
\usepackage{xspace}
\usepackage{cancel}

\usepackage{amssymb,url}
\usepackage{graphicx}
\usepackage{hyperref}

\usepackage{color}

\usepackage{array}
\usepackage{amsmath}
\usepackage{slashed}

\definecolor{rossoCP3}{cmyk}{0,.88,.77,.40}

\long\def\del #1 \enddel { }

\usepackage{graphicx}
\usepackage{amsmath}
\usepackage{amssymb}
\usepackage{subfigure}

\usepackage{epsfig}

 \usepackage{float}

\usepackage{graphicx}
\usepackage{subfigure}
\usepackage{hyperref}

\def\beq{\begin{equation}}
\def\eeq{\end{equation}}

\def\bea{\arraycolsep .1em \begin{eqnarray}}
\def\eea{\end{eqnarray}}
\def\Tr{{\rm Tr}}

\def\al#1{\alpha_{#1}}
\def\eq#1{(\ref{#1})}

\def\s0#1#2{\mbox{\small{$ \frac{#1}{#2} $}}}
\def\0#1#2{\frac{#1}{#2}}

\def\grgl{\:\hbox to -0.2pt{\lower2.5pt\hbox{$\sim$}\hss}{\raise3pt\hbox{$>$}}\:}
\def\klgl{\:\hbox to -0.2pt{\lower2.5pt\hbox{$\sim$}\hss}{\raise3pt\hbox{$<$}}\:}

\def\lsim{\mathrel{\rlap{\lower4pt\hbox{\hskip1pt$\sim$}}
    \raise1pt\hbox{$<$}}}                
\def\gsim{\mathrel{\rlap{\lower4pt\hbox{\hskip1pt$\sim$}}
    \raise1pt\hbox{$>$}}}                

\begin{document}
${}$\vskip1cm

\title{Conformal Gauge-Yukawa Theories away From Four Dimensions}
\author{Alessandro Codello}
\email{codello@cp3.sdu.dk}
\affiliation{{\color{rossoCP3}CP${}^3$-Origins},
	Univ. of Southern Denmark, Campusvej 55, DK-5230 Odense}
\author{Kasper Lang\ae ble}
\email{langaeble@cp3.sdu.dk}
\affiliation{{\color{rossoCP3}CP${}^3$-Origins},
	Univ. of Southern Denmark, Campusvej 55, DK-5230 Odense}
\author{Daniel~F.~Litim}
\email{d.litim@sussex.ac.uk}
\affiliation{\mbox{Department of Physics and Astronomy, U Sussex, Brighton, BN1 9QH, U.K.}}
\author{Francesco~Sannino}
\email{sannino@cp3.dias.sdu.dk}
\affiliation{{\color{rossoCP3}CP${}^3$-Origins},
	Univ. of Southern Denmark, Campusvej 55, DK-5230 Odense}
\affiliation{Danish Institute for Advanced Study {\color{rossoCP3}Danish IAS},
Univ. of Southern Denmark, Campusvej 55, DK-5230 Odense}

\begin{abstract}
\vskip0.5cm
We present the phase diagram and associated fixed points for a wide class of Gauge-Yukawa theories in  $d=4+\epsilon$ dimensions. The theories we investigate involve non-abelian gauge fields, fermions and scalars in the Veneziano-Witten limit.  The analysis is performed in steps, we start with QCD$_d$ and then we add Yukawa interactions and scalars which we study at next-to- and next-to-next-to-leading order. Interacting infrared fixed points naturally emerge in dimensions lower than four while ultraviolet ones appear above four. We also analyse the stability of the scalar potential for the discovered fixed points. 
We argue for a very rich phase diagram in three dimensions while in dimensions higher than four certain Gauge-Yukawa theories are ultraviolet complete because of the emergence of an  asymptotically safe fixed point. \\
\\ 
{\noindent \footnotesize Preprint: CP${}^3$-Origins-2016-12 DNRF90, DIAS-2016-12}
\end{abstract}

\maketitle

\section{Introduction}
There has been recent interest in the conformal structure of gauge theories in $d = 4+\epsilon$ dimensions with special attention to QED$_d$ \cite{Giombi:2015haa}. It has also been noticed that QED$_3$ can be viewed as the continuum limit of spin systems \cite{DiPietro:2015taa}. 

Of particular interest are  applications to phase transitions with multiple order parameters such as confinement and chiral symmetry breaking \cite{Sannino:2002wb,Mocsy:2003qw} at nonzero finite temperature and matter density. Analyses in $d = 4+\epsilon$ revealed intriguing possibilities such as the possible occurrence of tetracritical-type phase transitions \cite{Sannino:2004ix}. Related recent studies appeared in \cite{Eichhorn:2013zza,Eichhorn:2014asa,Eichhorn:2015woa}.  Away from four dimensions nonabelian gauge theories  have been studied in \cite{Peskin:1980ay,Appelquist:2000nn,Gies:2003ic,Morris:2004mg,Kazakov:2007su},
the non-linear sigma model in  \cite{Codello:2008qq} and scalar QED$_3$ in \cite{Freire:2000sx}. Similarly quantum gravity has been investigated in general dimensions in \cite{Litim:2003vp,Fischer:2006fz,Falls:2015qga}.

 Our goal is to go beyond the present knowledge by simultaneously extending the perturbative analysis beyond the leading order and by including gauge-singlet scalar degrees of freedom. The latter are in the form of complex scalar Higgs matrices that are  bi-fundamental with respect to the global symmetry group. In this initial investigation we work in the Veneziano-Witten limit  of infinite number of flavors and colors to neatly uncover the salient properties.  

In four dimensions a similar analysis has led to the discovery of the first nonsupersymmetric class of asymptotically safe quantum field theories \cite{Litim:2014uca}.  Furthermore the quantum corrected potential was determined in \cite{Litim:2015iea} while basic thermodynamic properties were uncovered in \cite{Rischke:2015mea}. The nonperturbative dynamics of  supersymmetric cousin theories in four dimensions was analysed in \cite{Intriligator:2015xxa}, further generalising or correcting the results of \cite{Martin:2000cr}. Here it was nonperturbatively  shown that the supersymmetric versions of the theories investigated in  \cite{Litim:2014uca} cannot be asymptotically safe. The discovery of asymptotically safe field theories in four dimensions has far reaching consequences for model building \cite{Sannino:2015fhk,Svendsen:2016kvn,Nielsen:2015una}.   

The work is organized as follows. We briefly introduce the Lagrangian for a {general class of Gauge-Yukawa theories in Section \ref{lagrangian}} together with the Veneziano-Witten rescaled couplings. { Here we also discuss  some general features of the $\beta$ functions of this class of theories in $d=4+\epsilon$ dimensions. In  Section \ref{QCD}, as a stepping stone we consider first the interesting case of QCD$_d$. We then upgrade it to the full Gauge-Yukawa system in Section \ref{GaugeYukawa}. We will consider the full Gauge-Yukawa system at both the next-to-leading order (NLO) and next-to-next-to-leading order (NNLO). We offer our conclusions and general considerations in Section \ref{Conclusion}.}
\section{Gauge-Yukawa Theory Template}
\label{lagrangian}

{ Our starting point is the general class of Gauge-Yukawa theories that, in four dimensions, constitutes the back-bone of the standard model of particle interactions. We consider an $SU(N_C)$ gauge theory featuring gauge fields $A_{\mu}^{a}$ with associated field strength $F_{\mu\nu}^{a}$, $N_F$ 
 Dirac  fermions $Q_i$ $(i=1, \cdots , N_F)$ transforming according to the fundamental representation of the gauge group (color-index muted), and an $N_F \times N_F$
complex matrix scalar field $H$ uncharged under the gauge group.
We take the fundamental action to be the same as in \cite{Antipin:2013pya,Litim:2014uca}, such that the Lagrangian is the sum of the following terms:}
\bea
\label{F2}
L_{\rm YM}&=& - \tfrac{1}{2} \Tr \,F^{\mu \nu} F_{\mu \nu}\nonumber\\
\label{F}
L_F\ &=& \ \Tr\left(
\overline{Q}\,  i\slashed{D}\, Q \right)
\nonumber\\
\label{Y}
L_Y \ &=&
\ y \, \Tr\left(\overline{Q}_L H Q_R + \overline{Q}_R H^\dagger Q_L\right)
\nonumber\\
\label{H}
L_H \ &=& \ \Tr\,(\partial_\mu H ^\dagger\, \partial^\mu H) \nonumber\\
\label{U}
L_U \ &=&
-u\,\Tr\,(H ^\dagger H )^2  \,\nonumber\\
\label{V}
L_V \ &=&
-v\,(\Tr\,H ^\dagger H )^2  \,. 
\eea
$\Tr$ is the trace over both color and flavor indices, and the decomposition $Q=Q_L+Q_R$ with $Q_{L/R}=\frac 12(1\pm \gamma_5)Q$ is understood. 
In four dimensions, the model has four classically marginal coupling constants given by the gauge coupling $g$, the Yukawa coupling $y$, the quartic scalar coupling $u$ and the `double-trace' scalar coupling $v$, which we write as:
\begin{align}
	\label{couplings}
	\al g & =\frac{g^2\,N_C}{(4\pi)^{d/2}\Gamma(d/2)}\,,\quad
	\al y=\frac{y^{2}\,N_C}{(4\pi)^{d/2}\Gamma(d/2)}\,,\quad\nonumber\\
	\al h & =\frac{{u}\,N_F}{(4\pi)^{d/2}\Gamma(d/2)}\,,\quad
	\al v=\frac{{v}\,N^2_F}{(4\pi)^{d/2}\Gamma(d/2)}\,.
\end{align}
We have normalized the couplings with the appropriate powers of $N_C$ and $N_F$ { so that the limit of infinite number of colors and flavours with $N_F/N_C$ a real finite number (Veneziano-Witten limit) is well defined. In dimensions different from four the couplings are dimensionful. We will denote by }
$\beta_i$ with $i=(g,y,h,v)$ the $\beta$-functions for the dimensionless version of the couplings in \eq{couplings}, which we will still call $\alpha_i$ for simplicity. {  We express the $\beta$ functions in terms of  
\begin{equation}\delta= \frac{N_F}{N_C}-\frac{11}{2} \ ,\end{equation} 
which in  Veneziano-Witten limit is a continuous parameter taking values in the interval $[-\tfrac{11}{2}, \infty)$.}

\begin{figure}[t]
	\begin{center}
		\includegraphics[width=0.8\linewidth]{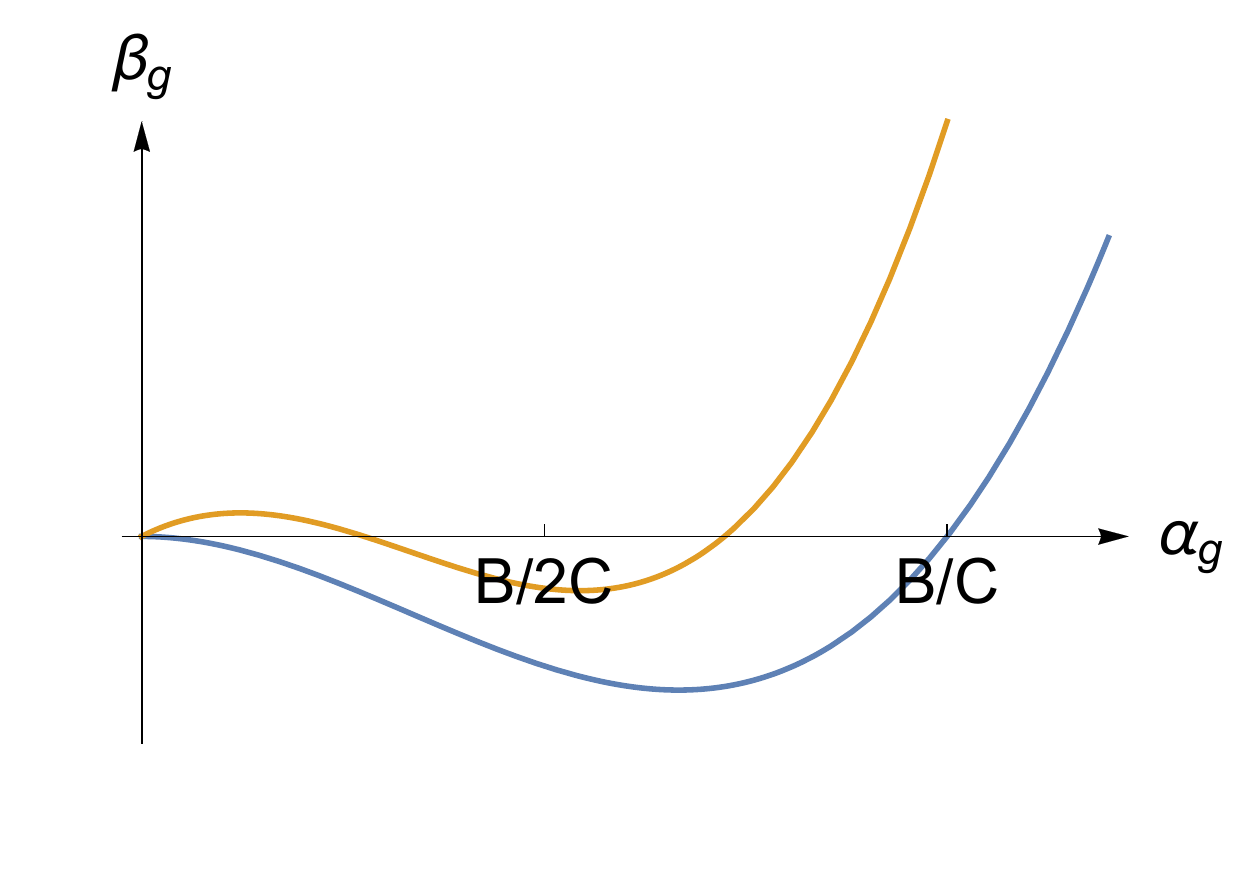}
		\caption{The blue curve corresponds to a four-dimensional effective gauge $\beta$-function of \eqref{GenericBetaQCD} that is asymptotically free (Gaussian UV fixed point) and further develops an interacting IR fixed point (i.e B,C $>0$). The orange curve displays the effects of introducing a small, but positive $\epsilon$ to four-dimensions, i.e.  moving towards higher dimensions. In this case, the Gaussian UV fixed point splits into a Gaussian IR fixed point and a interacting UV fixed point. \label{Fig-BetaCartoon}}
	\end{center}
\end{figure}

Since we perform the $\epsilon$-expansion around four dimensions, we require the $\beta$-functions to abide the Weyl consistency conditions \cite{Antipin:2013pya,Jack:1990eb,Osborn:1989td}.  This implies that, at the LO, one only needs to consider the one-loop corrections to the gauge coupling. We denote this order with  (1,0,0,0) to indicate zero loops in the other three couplings.  Therefore to LO the Gauge-Yukawa system reduces to the LO of QCD$_d$. We will also consider the phase diagram, beyond the LO, for QCD$_d$ in section \ref{QCD}. At the NLO the Gauge-Yukawa system (2,1,0,0) can be simplified by first setting the Yukawa $\beta$ function  to zero. This allows to write the Yukawa coupling, at the fixed point, as function of the gauge coupling. One can then substitute into the gauge $\beta$ function.  To this order we can anticipate the generic form of the associated fixed points by analysing the expected  two-loop gauge-beta function in terms of the gauge coupling. The situation is more involved at the NNLO.   At LO the gauge $\beta$-function  in $d=4+\epsilon$ does not depend on the Yukawa coupling and reads
\begin{align}\label{GLOBetaQCD}
\beta_g &=\epsilon\alpha_g-B\alpha_g^2\, \ . 
\end{align} 
In four dimensions, i.e. $\epsilon=0$, and to this order the theory displays only a non-interacting fixed point.  For $B>0$ it corresponds to asymptotic freedom and for $B<0$ to an infrared free theory. Away from four dimensions, i.e. $\epsilon\neq0$, we have a Gaussian fixed point (G) and a non-Gaussian fixed point (NG) given by
  \begin{equation}
\alpha^{G}_{g} =0 \ , \qquad 
\alpha^{\rm NG}_{g} =\frac{\epsilon}{B} \end{equation}  
Above four dimensions an ultraviolet safe fixed point emerges for $B>0$ and vice-versa for $B<0$ the interacting fixed point becomes an infrared one.  That means that an asymptotically free theory in four dimensions leads to an asymptotically safe one in higher dimensions while an infrared free leads to a theory with an interacting fixed point in dimensions lower than four. At the NLO, and upon having substituted the Yukawa coupling as function of the gauge coupling into the gauge beta function, the effective gauge $\beta$-function  reads
\begin{align}\label{GenericBetaQCD}
\beta^{\rm eff}_g &=\epsilon\alpha_g-B\alpha_g^2+C\alpha_g^3\, \ . 
\end{align} 
In four dimensions for $B>0$  one achieves an interacting IR fixed point for $\alpha_g^*=B/C$ with $C>0$. When asymptotic freedom is lost, $B<0$, the theory can have an asymptotically safe UV fixed point at $\alpha_g^*=B/C$ for $C<0$ \cite{Litim:2014uca}. Away from four dimensions, i.e. $\epsilon\neq0$, at the NLO we have at most three fixed points: a Gaussian fixed point (G) and two non-Gaussian fixed points (NG$_{\pm}$) given by
\begin{align}\label{GenericQcdFPs}
\alpha^{G}_{g} &=0\nonumber\\
\alpha^{\rm NG_{\pm}}_{g} &=\frac{B\pm\sqrt{\Delta}}{2C}=\frac{B}{2C}\left(1\pm\sqrt{1-4\epsilon\frac{C}{B^2}}\right),
\end{align}
where $\Delta=B^2-4 \epsilon C$.  In figure (\ref{Fig-BetaCartoon}) we show the $\beta$-function of (\ref{GenericBetaQCD}) with and without the linear term in $\alpha_g$.

\begin{table}
	\small\small
	\begin{tabular}{ccccccc}
		$d$ & $\alpha_{g}^{\rm G}$  & $\alpha_{g}^{\rm{NG}_{-}}$  & $\alpha_{g}^{\rm{NG}_{+}}$  & $\theta_{G}$ & $\theta_{NG_{-}}$ & $\theta_{NG_{+}}$ \\ 
		\hline $3$ & 0 & 0.0072 & -0.1567 & -1 & 1.0457 & 22.892 \\ 
		$5$ & 0 & 0.1033 &-0.4269 & 1 & -1.2421 & -5.1304  \\ 
		\hline 
		\hline
	\end{tabular}
	\caption{\label{TabQCD-NLO} Summary of the scaling exponents at NLO in QCD$_d$ for $d=3$, ($d=5$) assuming $\delta=100$, ($\delta=-5.5$)}
\end{table}
{ In $d=4$, perturbativity of the non-Gaussian (either IR or UV) fixed point is guaranteed for $\left|B\right|\ll1$ and  $\left|C\right|$  of order unity \cite{Banks:1981nn}. The situation changes when we go away from four-dimensions. By inspecting \eqref{GenericQcdFPs} we can consider different regimes. The first is the one in which $ \epsilon $ vanishes more rapidly than $|B|^2$ (proportional to $\delta^2$). This regime is a slight modification of the four dimensional case.  To be able to extend our analysis to finite values of $\epsilon$ (ideally achieving integer dimensions above and below four), we need $B^2>\left|C\right|$. For the case $B^2\gg\left|C\right|$ and $C$  of order unity, we have a perturbative fixed point $\alpha^{*}_{g} \simeq\epsilon/B$ and a non-perturbative one at $\alpha^{*}_{g} \simeq B/C$. Only when $B/C>0$ the second fixed point is physical. The first one is physical when $\epsilon$ and $B$ have the same sign. Also, a negative $\epsilon$ corresponds to having an UV fixed point for vanishing couplings. When both non-Gaussian fixed points are physical, they are separated by $\sqrt{\Delta}/C$. They will thus merge when $\Delta=0$. A near conformal behaviour is expected for very small negative values of $\Delta$ since the $\beta$-function almost crosses zero. This situation, however, is very different from the four-dimensional case (see \cite{Sannino:2009za} for a recent review) because: i) it arises away from four dimensions; ii) it can appear already at two-loop level while it requires at least three loops in four dimensions \cite{Pica:2010xq,Ryttov:2010iz}. In the case of one coupling, there is only one scaling exponent which is simply the derivative of the $\beta$-function at the fixed point, $\theta=\beta'(\alpha_g^*)$: 
\begin{align}\label{critexp}
\theta_{\rm G} = \epsilon\qquad \quad \theta_{\rm NG_{\pm}} =\frac{\Delta\pm B\sqrt{\Delta}}{2C}.
\end{align}
We will discuss the scaling exponents when we have explicit forms for B,C in the next sections.}

\section{QCD$_d$}\label{QCD}
Here we elucidate the phase diagram for QCD$_d$ before embarking on the phase diagram of the full Gauge-Yukawa theory in the next sections. 

\subsection{QCD$_d$ at leading order}
At the leading order we have%
\begin{align}\label{BetaLO}
\beta_g &=\epsilon\alpha_g+\frac43\delta\,\alpha_g^2\, .
\end{align}
The term linear in $\alpha_g$ and proportional to $\epsilon$ appears when rendering dimensionless the coupling in $d=4+\epsilon$ dimensions via  the replacement ${\alpha}_i \rightarrow \mu^\epsilon\alpha_i$ with $\mu$ the RG scale. {In addition to the simple zero at the origin $\alpha_g^{\rm G}=0$, the beta function of \eqref{BetaLO} has a non-trivial zero at }
\begin{equation}\label{LOfp}
\alpha^{\rm NG}_{g}=-\frac{3}{4}  \frac{\epsilon}{\delta}  \,.
\end{equation} 
{Positivity of $\alpha_g$ requires  that below (above) four  dimensions we must have $\delta>0~(\delta< 0 )$. In dimensions higher than four the non-interacting fixed point is IR while the interacting one is UV safe. In dimensions lower than four the UV and IR roles are inverted. 
Differently from the four-dimensional case perturbativity is guaranteed by a non-vanishing value of $\delta$ for any nonzero and small $\epsilon$.  }
Larger values of $\delta$, implying going away from the asymptotic freedom boundary in four-dimensions, naively seem to allow  larger values of $\epsilon$ that could potentially assume integer values\footnote{One could naively consider arbitrary large positive values of $\delta$  because we can have $N_f \gg N_c$. On the other hand  the smallest negative value of $\delta$ occurs for $N_f=0$ corresponding to $\delta =   -11/2$.}. This is, however, unsupported by a careful analysis of higher order corrections.  At each new order higher powers of $\delta$ appear. They are organised as shown in  \cite{Pica:2010xq,Gracey:1996he} and therefore in the infinite $\delta$ limit the couplings must be properly redefined. We shall not consider this limit here and we will  analyze our results for finite values of $\delta$. We will however further test  the stability of our results by comparing LO, NLO and NNLO approximations when assuming integer values of $\epsilon$.
The LO scaling exponents are  $\theta_{\rm G}=\epsilon$ and $\theta_{\rm NG}=-\epsilon$ for either signs of $\epsilon$, and do not depend on $\delta$. Clearly for positive $\epsilon$ the interacting fixed point is UV safe since the scaling exponent is negative indicating a relevant direction driving away from it.

\subsection{QCD$_d$ at next-to-leading order }

Adding the two loop contribution we arrive at the gauge $\beta$-function of the form (\ref{GenericBetaQCD})
\begin{align}\label{BetaPureGauge}
\beta_g &=\epsilon\alpha_g+\frac43\delta\, \alpha_g^2+\left(25+\frac{26}{3}\delta\right)\alpha_g^3.
\end{align}
To this order a third zero of the beta function appears of the form anticipated in \eqref{GenericQcdFPs} with $C=25+\frac{26}{3}\delta$.  The LO  $\alpha^{\rm NG}_g$ splits into $\alpha^{\rm NG_{+}}_g$ and $\alpha^{\rm NG_{-}}_g$.
 
Towards three dimensions, i.e. $ -1 \leq \epsilon<0$, we have that $\alpha^{\rm NG_{-}}_g$ naively vanishes for $\delta\rightarrow\infty$. When approaching five dimensions, i.e.  $1 \geq \epsilon>0$, one finds that $\alpha^{\rm NG_{-}}_g$ can be as small as  $\frac{1}{68} \left(\sqrt{204 \epsilon +121}-11\right)$ for the lowest admissible value of $\delta=-\tfrac{11}{2}$. Both fixed points vanish with $\epsilon$.  For $\delta \in [0,-\tfrac{75}{26}]$, both $B$ and $C$ are positive so that the interacting fixed points $\alpha^{\rm NG_{\pm}}_g$ are positive in dimensions higher than four, i.e. with $\epsilon>0$. Otherwise, B and C have opposite signs, and one of the two couplings is unphysical being negative. Furthermore by tuning $\delta$ the interacting fixed points disappear via a merging phenomenon.  The condition for this to happen is obtained by setting $\Delta=0$, or more explicitly
\begin{equation}
\left(\frac43\delta_c\right)^2=4 \epsilon \left(25+\frac{26}{3}\delta_c\right)   \, ,
\end{equation}
This yields a critical $\delta_c$. For $\epsilon=0$ we find $\delta_c = 0$, which is the point where asymptotic freedom is lost. In the general case, we find $\delta_c=\tfrac{3}{4} \left(13 \epsilon -\sqrt{\epsilon  (169 \epsilon +100)}\right)$. In $d=5$ the merger occurs for $\delta_c \simeq-2.551$ with $\alpha_g^{\rm NG_{\pm}}\simeq0.588$, which is a too large value for perturbation theory to hold.  

It is instructive to expand $\alpha_g^{\rm NG_{-}}$ in powers of $\epsilon$  \begin{equation}\label{QCDfpNLO}
\alpha_g^{\rm NG_{-}}=-\frac{3 \epsilon }{4 \delta }-\frac{9 (26 \delta +75)}{64 \delta}\left( \frac{\epsilon}{\delta}\right)^2 + \mathcal{O}(\epsilon^3) \, .
\end{equation}
In agreement with perturbation theory we find that the leading order term in $\epsilon$ matches the LO case, while corrections appear to order $\epsilon^2$. 

We report the NLO scaling dimensions  \eqref{critexp} in Table \ref*{TabQCD-NLO}.  The full four loop investigation in four dimensions of the phase diagram of QCD like theories has been performed in \cite{Pica:2010xq}.  It can serve as basis, in the future, to go beyond the NLO analysis away from four dimensions.

\begin{figure*}[t]
	\begin{center}
		\includegraphics[width=0.4\linewidth]{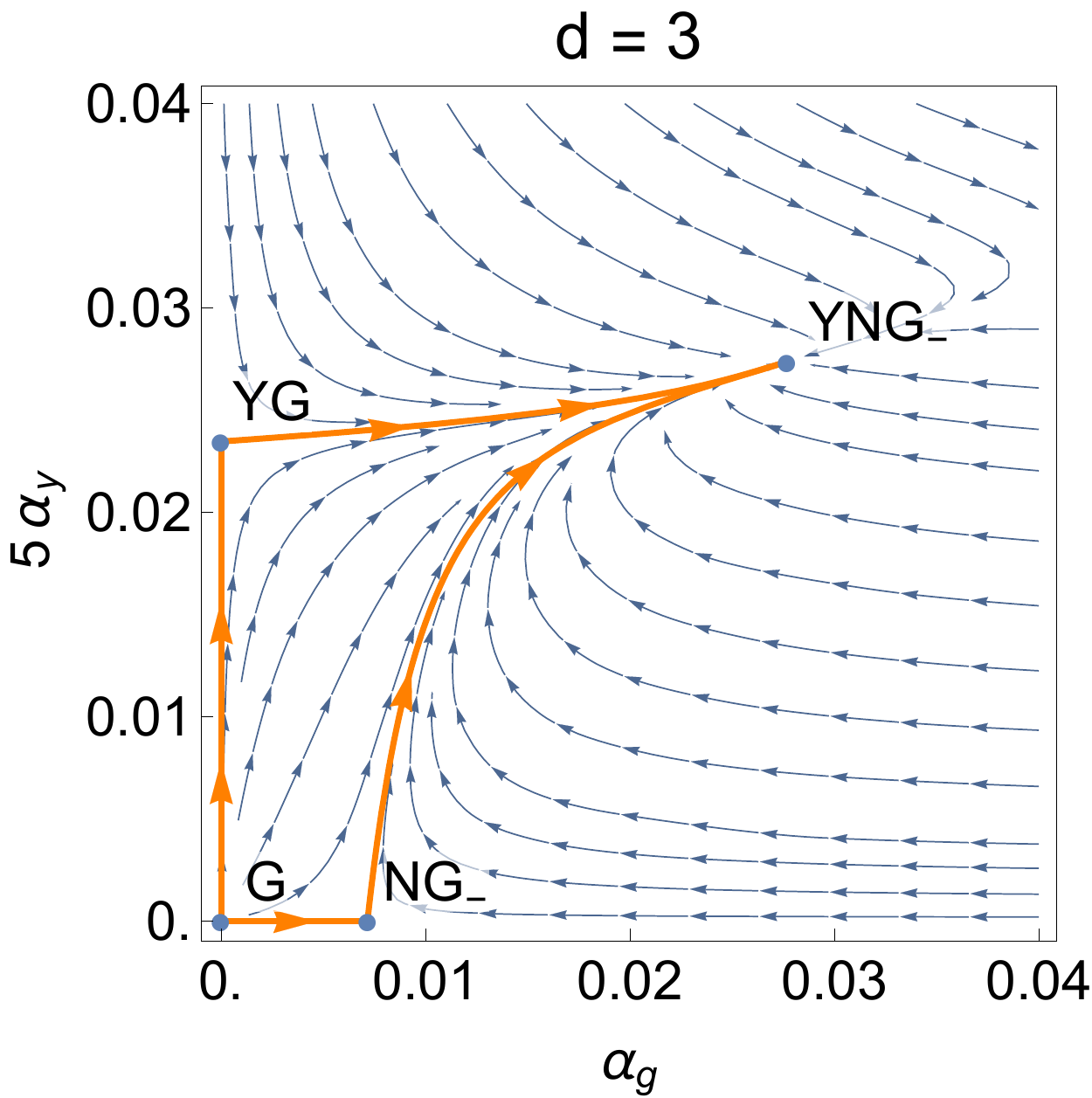}\qquad
		\includegraphics[width=0.4\linewidth]{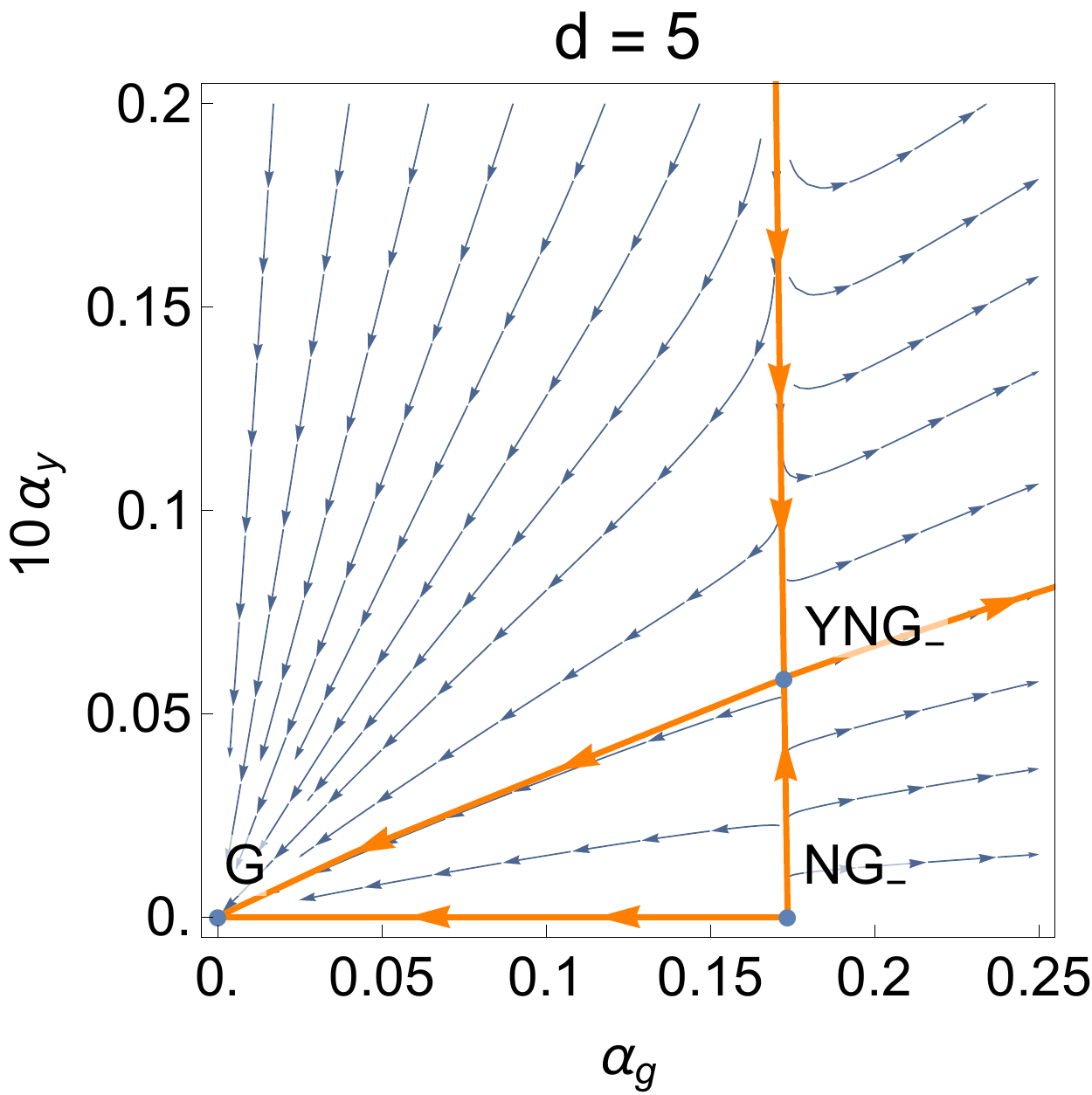}\
		\caption{The $(\alpha_{g},\alpha_{y})$-phase diagrams at NLO. \textit{Left}: NLO and $\epsilon=-1$ and $\delta=100$. \textit{Right}: NLO and $\epsilon=1$ and $\delta=\delta_*+0.1=-3.56$. \label{Fig-PhaseDiagramNLO}}
	\end{center}
\end{figure*}

\section{Gauge Yukawa}\label{GaugeYukawa}
We will now move to the full Gauge-Yukawa system. As already noticed, at LO it coincides with  LO-QCD$_d$. At the  NLO the Yukawa coupling starts playing an important role  \cite{Litim:2014uca}  affecting the  phase diagram of the theory.   The quartic couplings $\alpha_h$ and $\alpha_v$ will be relevant at the NNLO discussed in section \ref{beta functionsNNLO}. 
\subsection{Next-to-leading order}
The $\beta$-function system at NLO is \cite{Machacek:1983tz}
\begin{widetext}
\begin{align}\label{BetaNLO}
\beta_g &=\epsilon\alpha_g+\alpha_g^2\left\{\frac43\delta +\left(25+\frac{26}{3}\delta\right)\alpha_g-2\left(\frac{11}{2}+\delta\right)^2 \alpha_y \right\}\nonumber\\
\beta_y &=\epsilon\alpha_y+\alpha_y\left\{\left(13+2\delta\right)\alpha_y-6\alpha_g \right\}\, .
\end{align}
\end{widetext}
 To solve for the fixed points analytically it is convenient to  set to zero first the Yukawa beta function. We find two solutions: one for which $\alpha_y$ vanishes identically and that corresponds to the decoupled Yukawa limit with fixed points given by NLO-QCD$_d$; the second is the interacting one for which 
\begin{equation}\label{YukawaSol}
\alpha_y(\alpha_g)=\frac{6\alpha_g-\epsilon}{13+2\delta}\, .
\end{equation}
Upon substituting back the interacting solution in the gauge beta function it reduces to the form given in \eqref{GenericBetaQCD}. 

For $\alpha_g^{\rm G}=0$ we have the purely Yukawa interacting fixed point at $\alpha_y^{\rm YG}=\frac{-\epsilon}{13+2\delta}$. Since $\delta$ ranges in the interval $[-\tfrac{11}{2}, \infty)$ the denominator is always positive. This implies that  this fixed point  exists only in dimensions lower than four. 

Dimensions higher than four, i.e. $\epsilon>0$, can be achieved for $\alpha_g>\tfrac{\epsilon}{6}$. For $\epsilon<0$, the fixed point  satisfies $\alpha_y>\tfrac{\left| \epsilon \right|}{13+2\delta}$.

To be explicit, by inserting \eqref{YukawaSol} into \eqref{BetaNLO} we obtain the  effective gauge $\beta$--function:
%
\begin{align}\label{EffectiveBetaNLO}
\beta_g^ {\rm eff}=&\epsilon\alpha_g+\left\{\frac{4\delta}{3}+\frac{2\epsilon\left(\delta+\tfrac{11}{2}\right)^2}{2\delta+13}\right\}\alpha_g^2\nonumber\\
\qquad&+\frac{2\left(8\delta^2+46\delta-57\right)}{3(2\delta+13)}\alpha_g^3\,.
\end{align}
%
This equation assumes, of course, the same form as in \eqref{BetaPureGauge}. The three solutions are labelled by YG and YNG$_{\pm}$. As discussed above the YG fixed point exists only for $\epsilon < 0$.

In five dimensions, the gauge component of YNG$_{-}$ goes to $\frac{1}{68} \left(5 \sqrt{13}-11\right)\sim 0.1$ for $\delta\to-\frac{11}{2}$. Furthermore when the gauge coupling crosses ${\epsilon}/{6}$ the Yukawa coupling becomes negative. The crossing occurs for $\delta_*$, which at NLO reads
\begin{equation}\label{dcritNLO}
\delta_* = -\frac{3 (25 \epsilon +36)}{26 \epsilon +24} \, ,
\end{equation}
with $\delta_* =-3.66$ for $\epsilon=1$. We have therefore chosen the numerical value of $\delta=\delta_*+0.1$ for the phase diagram shown in Figure \ref{Fig-PhaseDiagramNLO}.
 For small $\epsilon$ we have  {\small
\begin{align}
\alpha_g^{\rm YNG_{-}}&=-\frac{3 \epsilon }{4 \delta }+\frac{9 \left(4 \delta ^3+36 \delta ^2+75 \delta +57\right)}{32 \delta (2 \delta +13)}\left( \frac{\epsilon}{\delta}\right)^2+\mathcal{O}(\epsilon^3)\nonumber\\
\alpha_y^{\rm YNG_{-}}&=\frac{ (-2 \delta -9)}{2   (2 \delta +13)}\frac{\epsilon}{\delta} +\mathcal{O}(\epsilon^2)\, .
\end{align} }
The equations simplify in the large $\delta$ limit
\begin{align}
\alpha_g^{\rm YNG_{-}}&=\left(-\frac{3}{4}+\frac{9 \epsilon}{16}\right)\frac{\epsilon}{\delta}+\frac{45}{32}\left( \frac{\epsilon}{\delta}\right)^2+\mathcal{O}(\epsilon^3)\nonumber\\
\alpha_y^{\rm YNG_{-}}&=-\frac12\frac{\epsilon}{\delta} +\mathcal{O}(\epsilon^2)\, .\label{NLOilargee}
\end{align}
It is  clear that the $\epsilon/\delta$ expansion is not supported at the NLO because of the emergence of the $\epsilon^2/\delta$ term.    %
\begin{table*}
	\begin{tabular}{c|cc|cccc}
		\multicolumn{7}{c}{$d=3$}\\
		\hline
		\hline 
		\multicolumn{1}{c}{ } & \multicolumn{2}{c}{NLO}& \multicolumn{4}{c}{NNLO} \\
		\hline FP & $\theta_{g}$ & $\theta_{y}$  & $\theta_{g}$ & $\theta_{y}$  & $\theta_{u}$ & $\theta_{v}$ \\ 
		\hline G & -1 & -1  & -1 & -1 & -1 & -1\\ 
		YG & -1 & 1 & -1 & 0.848 & -1.105 & -1.072 \\ 
		NG$_{-}$ & 1.046 & -1.043  & 1.017 & -1.027 & -1 & -1\\ 
		YNG$_{-}$ & 2.216 & 0.633 & 1.078 & 0.754 & -1.101 & -1.070 \\ 
		\hline
		\hline 
	\end{tabular}
	\caption{\label{TabD3} Summary of scaling exponents at NLO and NNLO with $\epsilon=-1$ and $\delta=100$. The name of the scaling exponents are meant to indicate the direction to which the stream line would flow in the deep infrared. All the scaling exponents go to $\pm 1$ at NLO in the limit $\delta\rightarrow\infty$. }
\end{table*}
\begin{table*}
	\begin{tabular}{c|cc|cccc}
		\multicolumn{7}{c}{$d=5$}\\
		\hline
		\hline 
		\multicolumn{1}{c}{}& \multicolumn{2}{c}{NLO}& \multicolumn{4}{c}{NNLO} \\
		\hline FP & $\theta_{g}$ & $\theta_{y}$  & $\theta_{g}$ & $\theta_{y}$  & $\theta_{u}$ & $\theta_{v}$ \\ 
		\hline G & 1 & 1 & 1  & 1 & 1 & 1\\  
		NG$_{-}$ & -1.176 & -0.041  & -1.380 & -0.028 & 1 & 1 \\ 
		YNG$_{-}$ & -1.180 & 0.041 & -1.381 & 0.028 & 1.026 & 1.026\\
		\hline
		\hline 
	\end{tabular}
	\caption{\label{TabD5} Summary of scaling exponents at NLO with $\epsilon=1$ and $\delta=-3.56$ and NNLO with $\epsilon=1$ and $\delta=-4.88$. The name of the scaling exponents are meant to indicate the direction to which the stream line would flow in the deep infrared. }
\end{table*}
\paragraph*{Phase diagram.}

To illustrate our results we now discuss the overall phase diagram for the limiting physical cases of five and three dimensions.  

In Fig.~\ref{Fig-PhaseDiagramNLO}  we present the NLO phase diagram for $d=5$.  For negative values of $\delta$ we observe three fixed points, the Gaussian (G), the non-Gaussian (${\rm NG}_-$) and the Yukawa non-Gaussian (${\rm YNG}_-$) in the $(\alpha_g,\alpha_y)$-plane.   The first fixed point is IR attractive in both directions while the second is UV attractive in both directions. The third one has a relevant direction (with negative scaling exponent) and an irrelevant one. Along the relevant direction it constitutes an  asymptotically safe fixed point. This phase diagram is very similar to the one considered in \cite{Litim:2014uca} and points towards the possible existence of a fundamental five dimensional Gauge-Yukawa theory.

For the $d=3$ case, the purely Yukawa fixed point (YG) occurs on the $\alpha_y$ axis  but now G is an UV fixed point while ${\rm YNG}_-$ is an IR one. Both YG and ${\rm NG}_-$ are of mixed character with one of the two couplings vanishing. This is a safety-free situation similar to the one uncovered in four dimensions in \cite{Esbensen:2015cjw}.  In the IR they also both flow to the same theory, i.e. to the fixed point ${\rm YNG}_-$ which is attractive from both directions.  

For either  $d=3$ or $d=5$ the fixed points ${\rm NG}_+$ and ${\rm YNG}_+$ occur for negative $\alpha_g$ and are therefore unphysical.

\paragraph*{Critical exponents.}

The linear RG flow in the vicinity of a given fixed point assumes the form
\begin{equation}\label{beta}
\beta_i=\sum_j M
_{ij}\,(\alpha_j-\alpha_j^*)+{\rm subleading} \, ,
\end{equation}
where $i=(g,y)$ and $M_{ij}=\left.\partial\beta_i/\partial\alpha_j\right|_*$ is the stability matrix.
The eigenvalues of $M$ are universal numbers and characterize the scaling of couplings in the vicinity of the fixed point.
The scaling exponents in three and five dimensions are listed respectively in Table \ref{TabD3} and \ref{TabD5}. The values of the scaling exponents give quantitative meaning  to the phase diagrams we presented in the previous paragraph. \\
%
\begin{figure*}[t]
	\begin{center}
		\includegraphics[width=0.4\linewidth]{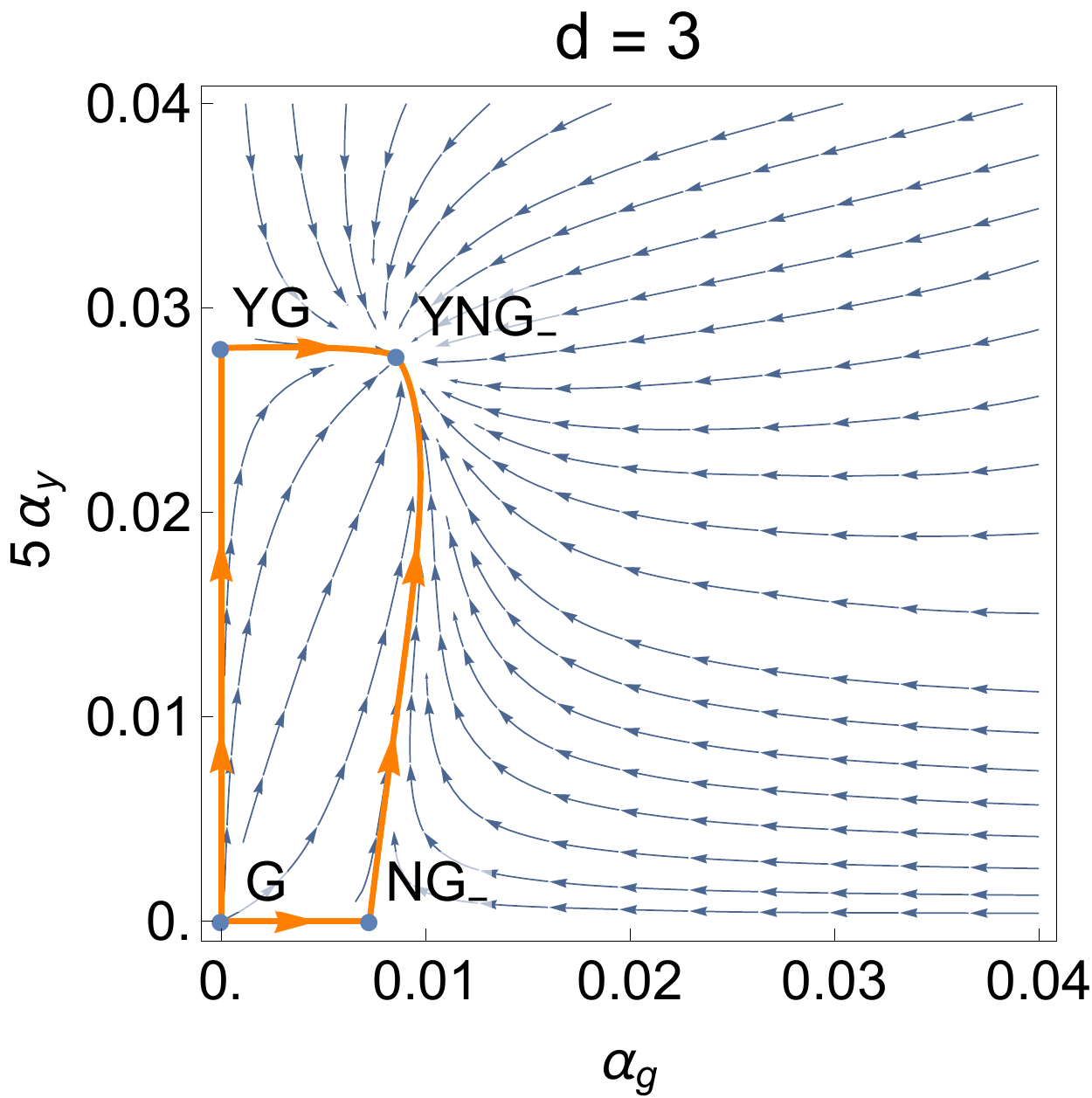}\qquad
		\includegraphics[width=0.4\linewidth]{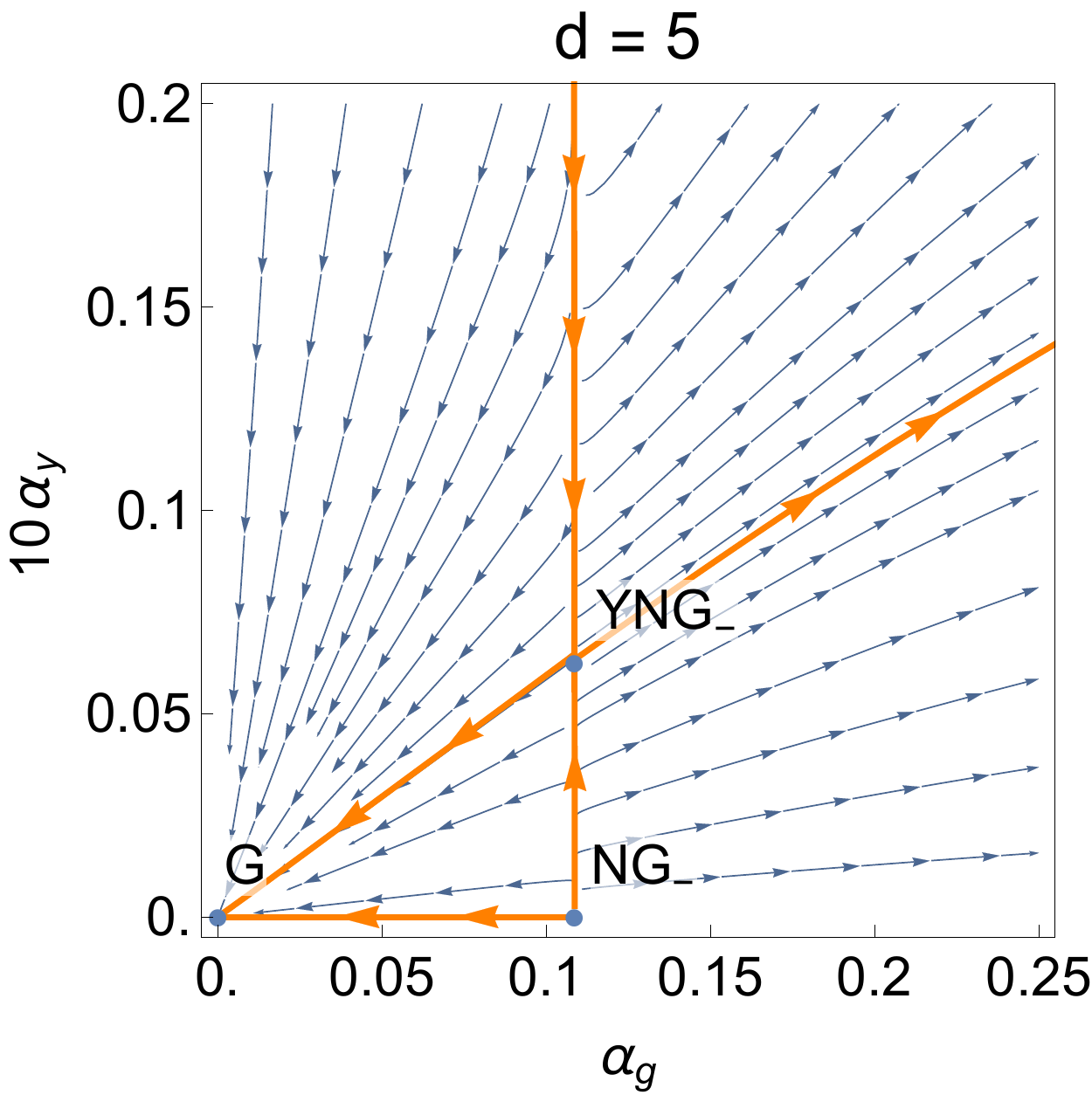}
		\caption{The $(\alpha_{g},\alpha_{y})$-phase diagrams at NNLO. \textit{Left}: NNLO and $\epsilon=-1$ and $\delta=100$. \textit{Right}: NNLO and $\epsilon=1$ and $\delta=-4.88$ \label{Fig-PhaseDiagramNNLO}}
	\end{center}
\end{figure*}
%
\subsection{Next-to-next-to-leading order}\label{beta functionsNNLO}
 In this section, we will consider the effect of adding NNLO-terms \cite{Machacek:1983tz}
\begin{widetext}
\begin{align}
\beta_g^{(3)} &=\alpha_g^2\left\{\left(\frac{701}{6}+\frac{53}{3}\delta-\frac{112}{27}\delta^2\right)\alpha_g^2-\frac{27}{8}(11+2\delta)^2\alpha_g\alpha_y+\frac14(11+2\delta)^2(20+3\delta)\alpha_y^2\right\}\nonumber\\
\beta_y^{(2)} &=\alpha_y\left\{\frac{20\delta-93}{6}\alpha_g^2+(49-8\delta)\alpha_g \alpha_y -\left(\frac{385}{8}+\frac{23}{2}\delta+\frac{\delta^2}{2}\right)\alpha_y^2-(44+8\delta)\alpha_y\alpha_h+4\alpha_h^2 \right\}\nonumber\\
\beta_h^{(1)} &=-(11+2\delta)\alpha_y^2+4\alpha_h(\alpha_y+2\alpha_h)\nonumber\\[3mm]
\beta_v^{(1)} &=12\alpha_h^2+4\alpha_v(\alpha_v+4\alpha_h+\alpha_y)\, .
\end{align}
\end{widetext}
\paragraph*{Phase diagrams and scaling exponents.} 

We now investigate the NNLO phase diagrams and explore the effects  of the scalar self-couplings. We will comment on the reliability of the NLO physical picture that emerged previously when assuming integer values of $\epsilon$. 
Figure \ref{Fig-PhaseDiagramNNLO} shows the NNLO version of the phase diagrams in three and five dimensions.
A comparison with their NLO counterparts, shown in Figure \ref{Fig-PhaseDiagramNLO},  manifest a certain stability:
in the $(\alpha_g,\alpha_y)$-plane the fixed points and the flow is qualitatively unchanged. 
But the new scalar directions affects the nature of the fixed points.
In $d=3$, as can be seen  from Table \ref{TabD3}, only G remains a complete UV-trivial fixed point while YG and NG$_-$ acquire two relevant directions in the scalar self-couplings.  Two relevant directions are added also to  YNG$_-$. We thus conclude  that in three dimensions  scalar self-couplings increase the dimension of the critical surface of YNG$_-$ and that the IR consequently reduces predictivity.

The $d=5$ dimension case is quite interesting since, as it is clear from Table \ref{TabD5},  both G and YNG$_-$ don't change their character: the first remains a complete IR fixed point while the second displays complete asymptotic safety with only one relevant direction. NG$_-$ adds two irrelevant directions to the previous two relevant ones.
 For the asymptotically safe  fixed point  YNG$_-$ one observes that the scaling exponents do not change much at the NNLO compared to the NLO case. The overall picture is that there is encouraging evidence for a five dimensional complete asymptotically safe Gauge-Yukawa theory. 

\begin{figure}[h!]
	\begin{center}
		\includegraphics[width=0.8\linewidth]{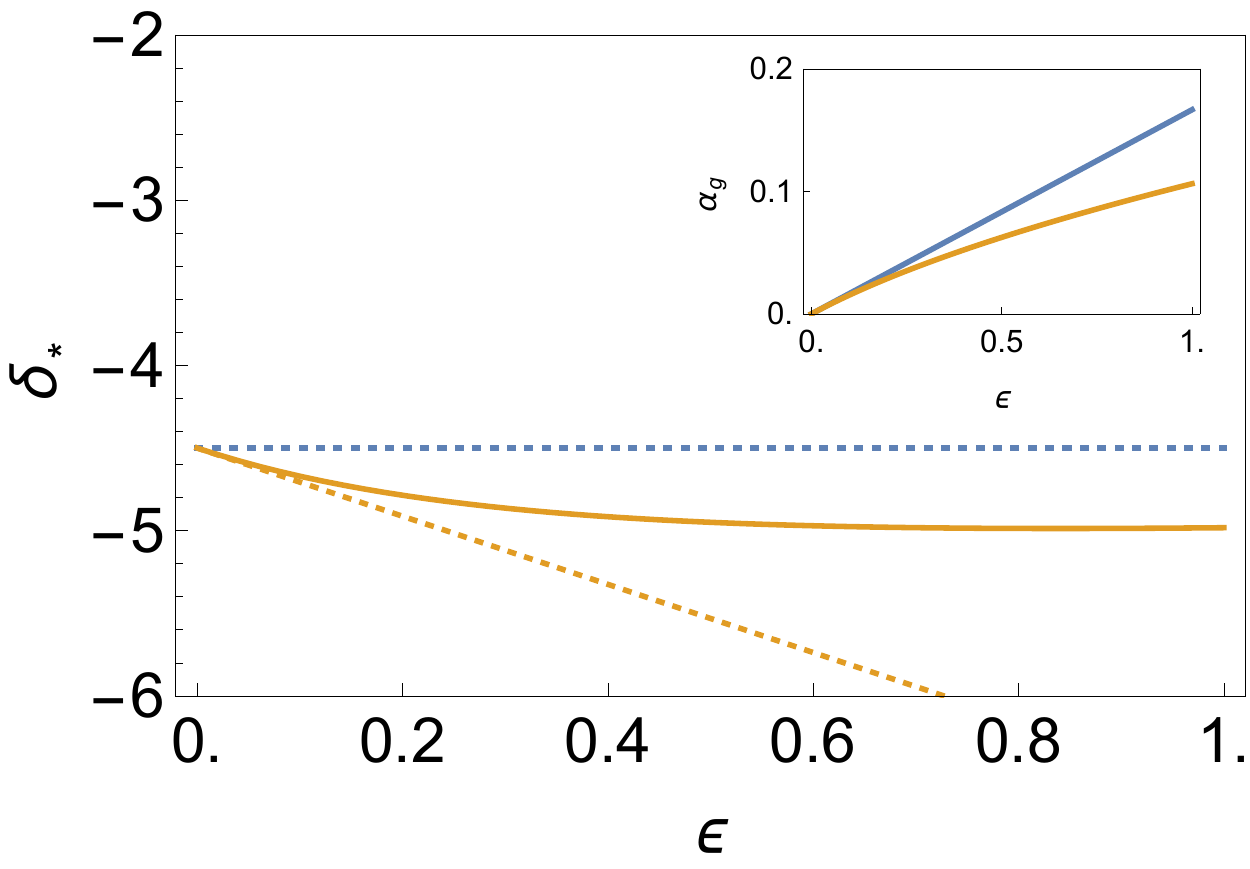}\
		\caption{Critical $\delta_*$ at NNLO (solid orange). The dotted lines are the constant (blue) and linear (orange) terms form an $\epsilon$-expansion of $\delta_*$. For a value of $\delta$ above the line, the theory has a fixed point with non-zero Yukawa coupling. The insert in the upper right corner shows the lower bound on $\alpha_g^*$ at NLO (\ref{YukawaSol}) in blue and NNLO (\ref{lowergNNLO}) in orange.  \label{fig:dcrit}}
	\end{center}
\end{figure}

\paragraph*{Stability of the scalar potential.} Here we will analyze the effect of adding loop contributions to the beta functions of the quartic couplings. In order for the scalar potential to be stable in the Veneziano-Witten limit, we need \cite{Litim:2015iea}
\begin{align}
\alpha_{h}^*>0&, \qquad \left(\alpha_{h}^*+\alpha_{v}^*\right)\geq 0\, . \label{quartic-constraints}
\end{align}
Solving the subsystem of the quartic couplings fulfilling this constraint, we find a required bound on the Yukawa coupling $\alpha_{y}\geq-\frac{\epsilon}{4}$. At the equality, we have $\alpha_{u}=-\alpha_{v}=\tfrac{\left|\epsilon\right|}{8}\sqrt{\tfrac{11}{2}+\delta}$. Hence, stability of the scalar potential puts a lower bound on the Yukawa coupling in dimensions lower than four. At the same time, it seems that in order to have small quartic couplings, we need $\delta$ negative. Combining this with the NLO Gauge-Yukawa system we predict for the pure Yukawa fixed point in three dimensions that $\delta$ is in the range $\left[-4.5,-5.5 \right]$ while the couplings are $\alpha_{g}=0,\alpha_{y}\sim 0.25,\alpha_{u}\sim-\alpha_{v}\sim 0.12$. For the interacting fixed point, the prediction is that $\delta$ is in the range $\left[1.5,4\right]$ while the couplings are $\alpha_{g}\sim 0.7,\alpha_{y}\sim 0.25,\alpha_{u}\sim-\alpha_{v}\sim 0.4$. We see that naively maintaining the condition for four dimensional stability of the scalar potential in three dimensions requires non-perturbative values of the fixed points.  Furthermore new superficially relevant operators can occur in three dimensions, which might change the stability condition \eqref{quartic-constraints}. We checked, via direct NNLO computation, that pushing for large negative values of $\epsilon$ is challenging from the scalar potential stability point of view.  In five dimensions, as we shall see, the system is less constrained resulting into a  stable scalar potential at the asymptotically safe fixed point. 
\paragraph*{The critical $\delta_*$}
In the analysis at NLO, we found that the gauge coupling crosses its lower bound $\alpha_g \geq \epsilon/6$ at the critical value $\delta_*$ (\ref{dcritNLO}). For the NNLO system this lower bound has changed into
\begin{equation}\label{lowergNNLO}
\alpha_{g}^*=\frac{18-\sqrt{6} \sqrt{-20 \epsilon  \delta +93 \epsilon +54}}{20 \delta -93}\, .
\end{equation}
If we solve the gauge beta function with Yukawa and quartic couplings set to zero, we can find a relation in $\delta(\epsilon)$ for which the equality is exactly true. This curve is shown on Figure \ref{fig:dcrit} together with the corresponding lower bound on $\alpha_g^*$ (\ref{lowergNNLO}). The critical value at NNLO evaluates to $\delta_*=-4.98$ at $\epsilon=1$. In the phase diagram shown in Figure \ref{Fig-PhaseDiagramNNLO} we use $\delta=\delta_*+0.1=-4.88$. 
We can now set up an expansion around $\delta_*$. In this way, we will develop a fully interacting asymptotically safe fixed point YNG$_-$ in 5D with perturbative control for the three couplings ($\alpha_y,\alpha_u,\alpha_v$). For a theory with $\delta=\delta_*+x$, where $x\ll1$, the Yukawa coupling emerges as linear in $x$, while the quartic couplings ($\alpha_u,\alpha_v$) are sub-leading, $x^2$ and $x^4$. The coefficient of $\alpha_u$ is positive, while it is negative for $\alpha_v$. Hence, the constraints (\ref{quartic-constraints}) are satisfied.\\

\section{Conclusion}\label{Conclusion}
 In this paper we have investigated Gauge-Yukawa theories in $d=4+\epsilon$ dimensions within the Veneziano-Witten limit and in perturbation theory.

 In $d=5$ the consistency between the NLO and NNLO phase diagrams points to the existence of an asymptotically safe fixed point. This result 
extends the discovery of asymptotic safe Gauge-Yukawa theories beyond four dimensions \cite{Litim:2014uca}. 
 
In three dimensions we found a rich phase diagram featuring, besides the UV non-interacting fixed point, several fixed points with the fully interacting one attractive in two directions but repulsive in the other two, rendering the theory less predictive in the deep infrared. 

Our analysis constitutes a step forward towards a systematic study of the phase diagram of Gauge-Yukawa theories in several space-time dimensions.

\acknowledgements 
 The CP${}^3$-Origins centre is partially funded by the Danish National Research Foundation, grant number DNRF90. This work is also supported by the Science Technology and Facilities Council (STFC) [grant number ST/L000504/1], by the National Science Foundation under Grant No.~PHYS-1066293, and by the hospitality of the Aspen Center for Physics.

\end{document}